\documentclass[10pt,conference]{IEEEtran} 
\IEEEoverridecommandlockouts
\usepackage{cite}
\usepackage{amsmath,amssymb,amsfonts}
\usepackage{algorithmic}
\usepackage{graphicx}
\usepackage{textcomp}
\usepackage{xcolor}
\usepackage{flushend}
\usepackage[breaklinks]{hyperref}
\usepackage{listings}
\usepackage[T1]{fontenc}

\def\BibTeX{{\rm B\kern-.05em{\sc i\kern-.025em b}\kern-.08em
		T\kern-.1667em\lower.7ex\hbox{E}\kern-.125emX}}

\begin{document}

\title{Detecting Exploit Primitives Automatically for Heap Vulnerabilities on Binary Programs}

\author{
	\IEEEauthorblockN{Jie Liu \IEEEauthorrefmark{1},  Hang An \IEEEauthorrefmark{2}, Jin Li\IEEEauthorrefmark{3}, Hongliang Liang\IEEEauthorrefmark{1}} 
	
	\IEEEauthorblockA{\IEEEauthorrefmark{1}School of Computer ScienceBeijing University of Posts and TelecommunicationsBeijing, Beijing, China} 
	\IEEEauthorblockA{\IEEEauthorrefmark{2}Codesafe Team of LegendsecQi’anxin Group, Beijing, China} 
	\IEEEauthorblockA{\IEEEauthorrefmark{3}Nation Key Laboratory of Science and Technology on Information System Security,  Beijing, China} 
	 	{\IEEEauthorrefmark{1}\{liujie\_ran, hliang\}@bupt.edu.cn,
	 		 \IEEEauthorrefmark{2}anhang@qianxin.com,  \IEEEauthorrefmark{3}tianyi198012@163.com}
}

\maketitle              
\begin{abstract}
Automated Exploit Generation (AEG) is a well-known difficult task, especially for heap vulnerabilities. 
Previous works first detected heap vulnerabilities and then searched for exploitable states by using symbolic execution and fuzzing techniques on binary programs.
However, it is not always easy to discovery bugs using fuzzing or symbolic technologies and solvable for internal overflow of heap objects.

In this paper, we present a solution DEPA to detect exploit primitives 
based on primitive-crucial-behavior model for heap vulnerabilities. 
The core of DEPA contains two novel techniques,
1) primitive-crucial-behavior identification through pointer dependence analysis,
and 2) exploit primitive determination method which includes triggering both vulnerabilities and exploit primitives.
We evaluate DEPA on eleven real-world CTF(capture the flag) programs with heap vulnerabilities and DEPA can discovery arbitrary write and arbitrary jump exploit primitives for ten programs except for program $multiheap$.
Results showed that  primitive-crucial-behavior identification and determining exploit primitives are  accurate and effective by using our approach. 
In addition, DEPA is superior to the state-of-the-art tools in determining exploit primitives for the heap object internal overflow vulnerability. 

\end{abstract}


\section{Introduction}

\begin{figure}[h]
	\centering
	\includegraphics[width=\linewidth]{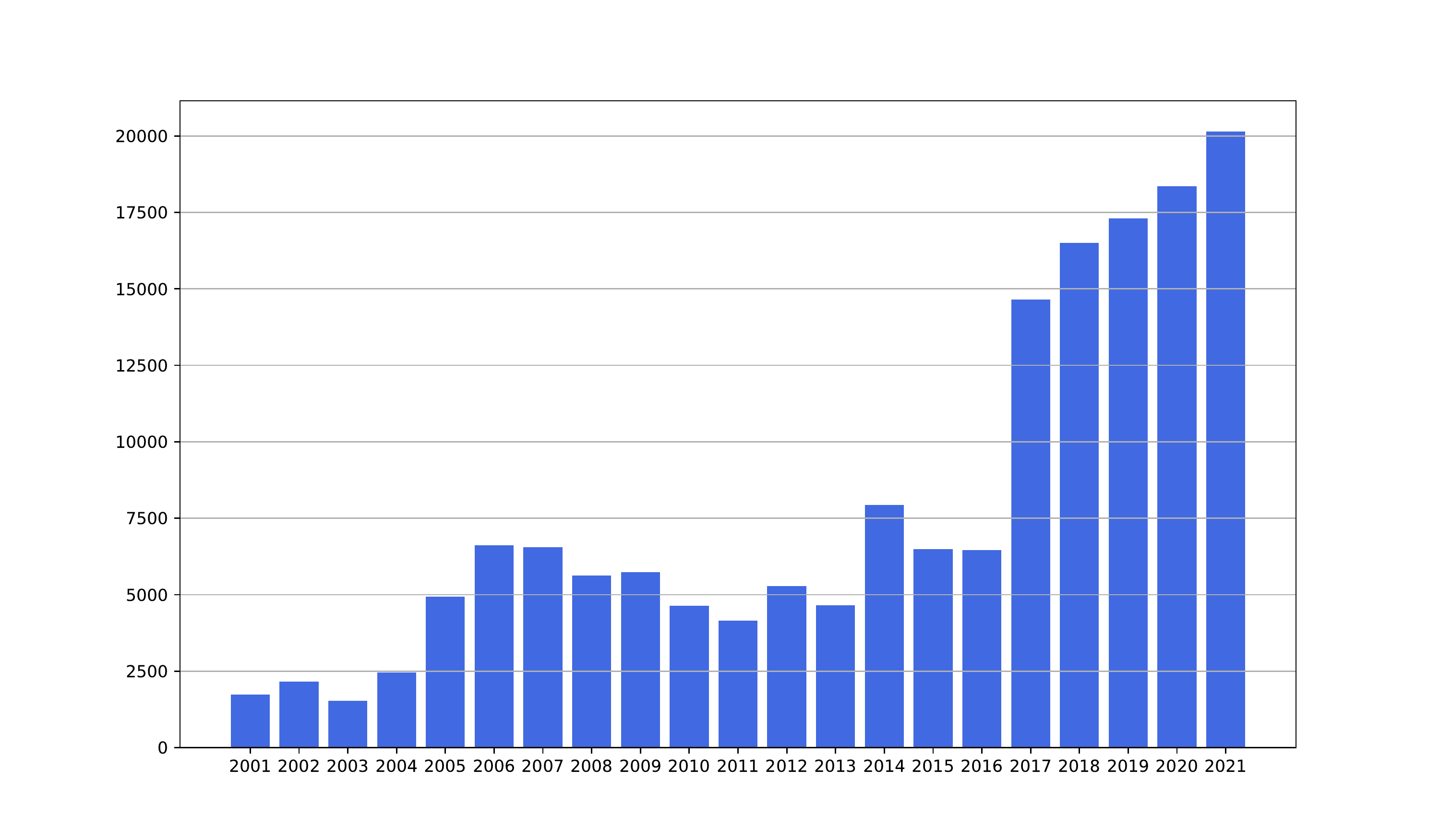}
	\caption{The distribution of vulnerabilities in NVD from 2001 to 2021 year}
	\label{NVDdata}
\end{figure}

In recent years, with the progress of automatic vulnerability mining technology, a number of vulnerabilities in various software are detected every year \cite{NVDdata}. 
According to statistics from the National Vulnerability Database (NVD) (Figure \ref{NVDdata}), the total number of vulnerabilities discovered in the last 5 years  is greater than  that  from 2001 to 2016, and the overall trend has been on the rise steeply since 2001.

With lots of bugs to fix, determining their severity is the first issue for security practitioners or software vendors. Security researchers also spend much time reproducing bugs, because the quality of bug reports is patchy in most cases. An effective way to determine the authenticity and severity of a vulnerability is to construct a working exploit. Unfortunately, it is usually a time-consuming and laborious work to construct such an exploit.  Therefore, the automatic exploit generation \cite{APEG,AEG,CRAX,HEELAN,VISH}  is a research topic in recent years.

Exploit primitives (or exploitable states) refer to program states that violate security requirements and enable attackers to obtain capabilities beyond those provided by the original program \cite{KEPLER}. 
In general, the determination of  exploit primitives is not only a prerequisite  but also the most challenging part for exploit generation. 
In this paper,  we therefore focus on the techniques of detecting exploit primitives, and take two typical exploit primitives as examples, i.e., arbitrary address writing and arbitrary address jump primitives.

Prior efforts  \cite{Q,CRAX,REVERY} in exploit primitive determination are  based on the capability of a  vulnerability through a POC input, which would crash the program. 
However, vulnerabilities for some event loop driven programs such as network interaction programs are hard to discover, 
and thus  they are not suitable to interaction programs for detecting exploit primitives.
We illustrate the challenges encountered during  detecting exploit primitives in an interactive program from CTF competition through the following code \ref{iprog}.
It accepts inputs from a user and deals with them accordingly.

\begin{figure}[h]
\begin{minipage}{0.5\textwidth}
\begin{lstlisting}[title=a) An example program (iProg)]
main(){
  short option;
  init();
  while(1){
    scanf("%d",&option);
    switch(option){
      case 1: writeNote();break;
      case 2: editNote();break;
      case 3: deleteNote();break;
      case 4: showNote();break;
      case 5: quit();break;
    }
  }
}
\end{lstlisting}
\end{minipage}
\begin{minipage}{0.5\textwidth}
\begin{lstlisting}[title=b) Structure Note in iProg]
struct Note {
  long Canary;
  int Index;
  int InUse;
  char Content[128];
  Note* Next; 
  Note* Prev; 
};
\end{lstlisting}
\end{minipage}
\caption{An example program and a structure in it.}
\label{iprog}
\end{figure}

In the program, the $ init()  $ function first allocates and initializes 20 heap objects (i.e., $Note$),  then one of five functions is executed according to a user input, i.e., $option$, and more user inputs are required to complete their functionalities respectively. For example,  the number of the edited note and the string of the edited note are further required inputs in $ editNote $.
In this paper, we call  such an operation, which is performed once through the main loop and further requires a sequence of  inputs according to a specific rule, as a meta-operation.

The following code (Listing \ref{editnote}) shows the $editNote()$ function, which contains an internal heap overflow vulnerability that will spill 8 bytes of  user input into the $ Next $ field in the struct $Note$ through $callRead()$ function at line 11.  The $ deleteNote()  $ function is shown as follows (Listing \ref{detelenote}), which dereferences the $ Next $ pointer and perform a write operation that triggers an arbitrary address write exploit primitive at line 6. 

The following code (Listing \ref{editnote}) shows the $editNote()$ function, which contains an internal heap overflow vulnerability that will spill 8 bytes of  user input into the $ Next $ field in the struct $Note$ through $callRead()$ function at line 11.  The $ deleteNote()  $ function is shown as follows (Listing \ref{detelenote}), which dereferences the $ Next $ pointer and perform a write operation that triggers an arbitrary address write exploit primitive at line 6.

\begin{lstlisting}[label=editnote, caption=editNote function]
int editNote(Note *headNote) {
  int index; 
  Note *selectedNote; 
  index = read_number();
  ...
  for (selectedNote = headNote; selectedNote && 
         selectedNote -> Index != index; 
         selectedNote = selectedNote->Next );
    if ( selectedNote && selectedNote -> InUse ) {
        // Overwrite of Next possible
        callRead(selectedNote -> Content, 136LL, 10);    
    }
  ...
}
\end{lstlisting}

\begin{lstlisting}[label=detelenote, caption=deleteNote function]
int deleteNote(Note *headNote) { 
  ...
      if ( selectedNote->InUse )
      {
        Note* ptr_prev_note = selectedNote->Prev;
        Note* ptr_next_note = selectedNote->Next;     
        // Unsafe unlink
        if ( ptr_prev_note )
          ptr_prev_note->Next = ptr_next_note;
        if ( ptr_next_note )
          ptr_next_note->Prev = ptr_prev_note;
        ...
      }
  ...
}
\end{lstlisting}

For the event loop driven program in the example,
how to determine exploit primitives for heap vulnerabilities is a  challenging task. 
And  the previous approaches usually continue to exploring the crashing paths in order to find an exploit primitive when discover vulnerabilities.
However, existing methods for discovering vulnerabilities, e.g. symbolic execution or fuzzing, can't deal well with the meta-operations for interaction programs 
and result in an inability to explore the deep states of the program.

Moreover, the vulnerability of internal overflow in heap objects is difficult to be identified \cite{REVERY}, which further increases the difficulty of detecting exploit primitives.
The process of identifying heap vulnerabilities is notoriously complex and time consuming. Unlike existing methods of detecting exploit primitives, our approach is to determine exploit primitives directly without detecting vulnerabilities.

We observe that triggering an exploit primitive for a vulnerability is closely related to some program semantics (functions or APIs) which can lead to or expose the vulnerability to some extent.
For instance, an exploit primitive for a heap vulnerability often requires a program path that goes through heap related semantics, e.g., heap object allocation, heap object write and heap object pointer dereference. 
As a result, we call such a function in a program as a primitive-crucial-behavior (PCB for shot).

Based on the above observation, we explore and leverage primitive-crucial-behaviors (PCBs) to guide the determination of exploit primitives, and we are facing with two challenges: 
1) How to identify the PCBs accurately? 
2) How can PCBs be applied to detect exploit primitives?

To solve these challenges, we propose a solution DEPA,
which could detect exploit primitives based on the analysis of PCBs.
First, DEPA identifies the PCBs corresponding to different types of inputs through static analysis technology,  
then constructs an input generation template according to the mapping from input type to PCB.
Next,  DEPA conducts fuzzing with the template on the target program, 
and finally  concolic execution is performed to assess whether the crash obtained 
by fuzzing is a working exploit primitive.

To sum up, the main contributions of our work are as follows:
\begin{enumerate}
\item We propose a method of program primitive-crucial-behaviors analysis, 
which is used to detect crucial behaviors of the program during symbolic execution; 
\item We devise an efficient method to determine exploit primitives by combining template-based fuzzing and concolic execution;
\item We implement an automated solution DEPA to detect exploit primitives and evaluation results show that DEPA is more effective than the state-of-the-state tool Revery in determining exploit primitive for the heap object internal overflow vulnerability.
\end{enumerate}

\begin{figure*}[t]
	\centering
	\includegraphics[width=0.6\textwidth]{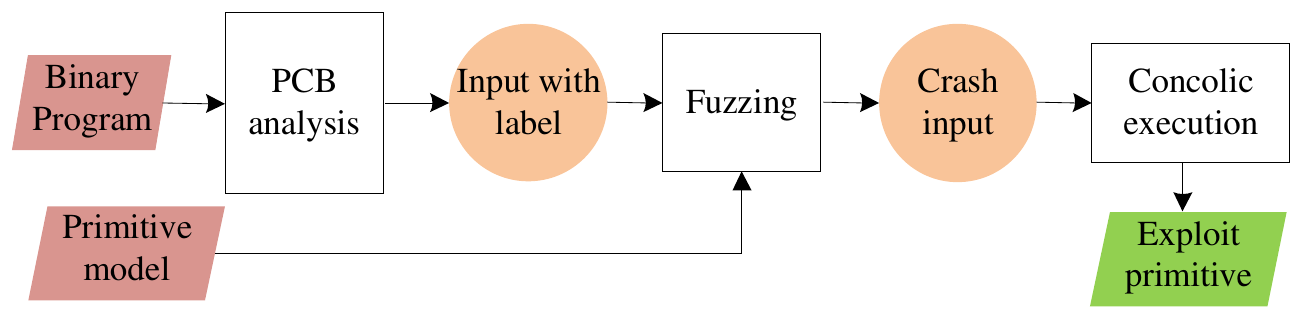}
	\caption{The architecture of DEPA}
	\label{architecture}
\end{figure*}

\begin{figure}[t]
	\centering
	\includegraphics[width=0.8\linewidth]{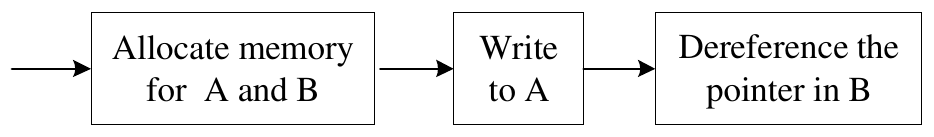}
	\caption{The process of triggering a heap overflow vulnerability exploit primitive}
	\label{OOB_process}
\end{figure}

\begin{figure*}[t]
	\centering
	\includegraphics[width=.6\linewidth]{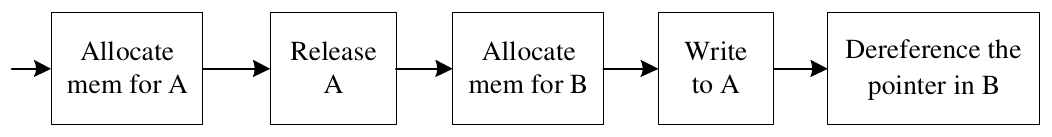}
	\caption{The process of triggering a UAF vulnerability exploit primitive}
	\label{UAF_process}
\end{figure*}

\begin{figure*}[t]
	\centering
	\includegraphics[width=.6\linewidth]{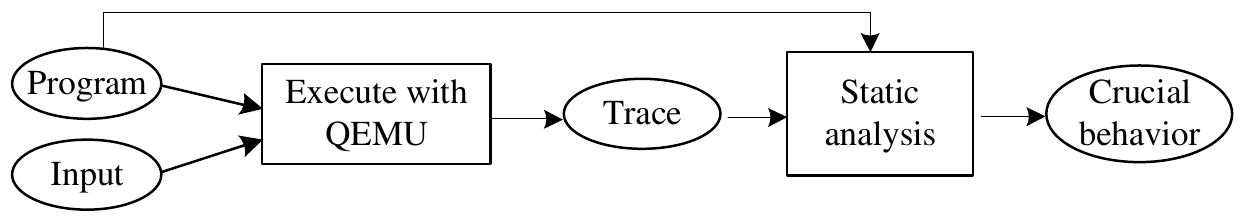}
	\caption{Primitive-crucial-behavior analysis}
	\label{behavior_analysis}
\end{figure*}

\section{Approach}

In this section, we describe the overall architecture of the DEPA system. 
As shown in Figure \ref{architecture}, the DEPA system contains three modules: 
(1) primitive-crucial-behavior analysis module, 
(2) fuzzing module based on generation, 
and (3) concolic execution module.
First, primitive-crucial-behavior analysis module takes target program as input and performs primitive-crucial-behavior analysis while symbolic executing the program. 
Different types of inputs will be labeled according to the corresponding primitive-crucial-behaviors.
Second, the input generation template will be constructed with the different exploit primitive model and tagged input. 
Using fuzzing based on the input generation template further gets some crash inputs that may trigger the exploit primitive.
Finally, the concolic execution module performs symbolic execution along to the crash path, 
and determines whether the crash input triggers the exploit primitive.

\subsection{Primitive-crucial-behavior analysis}
Given a binary program with vulnerability, DEPA first attempts to execute symbolically it 
and analyses primitive-crucial-behavior in order to attain some inputs with label.
Note that these inputs are kind and cannot crash the program. 
This includes two phases: input generation and primitive-crucial-behavior analysis.

\subsubsection{Input generation}\label{inputgen}
The idea of legal input generation module is direct.
The target program is executed symbolically, and then the corresponding constraint conditions are solved to get the input.
However, in the process of actual implementation, it is faced with the challenge of state explosion caused by loops inevitably because of the use of symbolic execution technique.

Existing symbol execution tools (e.g., angr) deal with loops by recording the number of times the loop executed 
and pausing the state if the number of loops exceeds a given threshold. 
This strategy is obviously crude, since the loops in a program can't always cause state explosions. 
If we blindly prune the loop, the number of paths that symbol execution can explore may decrease 
so that some crash paths hidden deep in the program may be missed.

For this reason, we classify the loops, and formulate different loop restriction strategies 
according to different types of loops, which can avoid the state explosion problem and ensure higher code coverage. 
Loops can be divided into the following three categories based on their escape conditions: 

\begin{itemize}
\item [a)] The case where the loop control expression is always True, e.g., \texttt{while(1)};

\item [b)] The case where the loop control expression is relative to the input, e.g., \texttt{while(input)};

\item [c)] The case where the value of the loop control expression is independent of the input but has a finite number of loops, e.g., \texttt{for(i =0; i$<$10; i++)}.
\end{itemize}

Type a) and b) both lead to state explosion, while type c) does not.
DEPA exploits different loop restriction methods to execute the loop of target program symbolically: 
\begin{itemize}
\item [a)] If the number of cycles of the current state exceeds the given threshold, the execution of the current state is suspended; otherwise, the execution of the current state continues.
\item [b)] If the break condition of the loop can be met, the state that breaks out of the loop will continue to execute, another state that continues the loop will suspend execution, otherwise the loop will continue. This avoids the state explosion caused by symbolic jump-out conditions.
\item [c)] No restrictions are placed.
\end{itemize}

\subsubsection{primitive-crucial-behavior analysis}

In this paper, heap vulnerability exploit primitive crucial-behavior is abbreviated as primitive-crucial-behavior.
The inputs of primitive-crucial-behavior analysis module are binary program and different types of input,
and the output is different types of input with labels. 
The labels represent the primitive-crucial-behavior of the program that the input can trigger. 
We define primitive-crucial-behavior as the program behavior that must exist on the program execution path to exploit primitive. 
This is the relationship between the triggering process of exploit primitives and the primitive-crucial-behavior.

Figure \ref{OOB_process} shows the common process of triggering a heap overflow vulnerability \cite{OOBdef} exploit primitive. 
First, adjacent objects A and B are allocated, 
and then if a heap overflow vulnerability occurs when writing object A,  
the spilled data can cover the pointer in object B. 
Finally, the pointer in object B is dereferenced so that the exploit primitive is triggered.

Figure \ref{UAF_process} shows the process of triggering a UAF \cite{UAFdef} vulnerability exploit primitive. 
Firstly, object A is freed and its pointer  does not set  NULL after it has been allocated.
Then object B whose size is approximately equal to object A is allocated and takes the memory allocated to object A.
Next the write operation to object A will overwrite the data in the object B.
If pointer in object B is overwritten, the dereference of this pointer in B will trigger the exploit primitive.

As can be seen from the triggering process of the two heap vulnerability exploit primitives mentioned above, 
the triggering process of exploit primitives depends on several primitive-crucial-behaviors of the program,
including object allocation, object write, object release, and object pointer dereference. 
We analyze the primitive-crucial-behaviors of the input to establish the relationship between the different types of input and the primitive-crucial-behaviors. 
The overall process of primitive-crucial-behavior analysis is shown in Figure \ref{behavior_analysis}.
The $ Input $ is first generated for target programs in input generation phase \ref{inputgen}.
Next, primitive-crucial-behavior analysis records the execution path of a particular input through concrete execution, 
and then captures the basic block information in the path.
Finally, the primitive-crucial-behavior is obtained by symbolic execution of the basic block.

The most important step in this process is how to analyze the binary code in the path 
and identify the primitive-crucial-behavior. 
We elevate the binary code to the intermediate representation (IR) code 
(e.g., VEX IR\footnote{https://docs.angr.io/advanced-topics/ir})
and analyze special behaviors based in the IR.
We identify the allocation and release of a heap object by judging 
whether the relevant functions (e.g., malloc, free) are called. 
Besides, the write to an allocated heap object is identified by using 
its pointer and the related instruction, e.g., Load instruction.
However, recognizing the dereference of a pointer in the heap object (i.e., in-object pointer) is not trivial, 
because this requires the point-to relationship between the heap object pointer and the in-object pointer. 
Therefore, we construct a pointer dependence analysis module based on the IR. 

\begin{figure}[h]
	\centering
	\includegraphics[width=\linewidth]{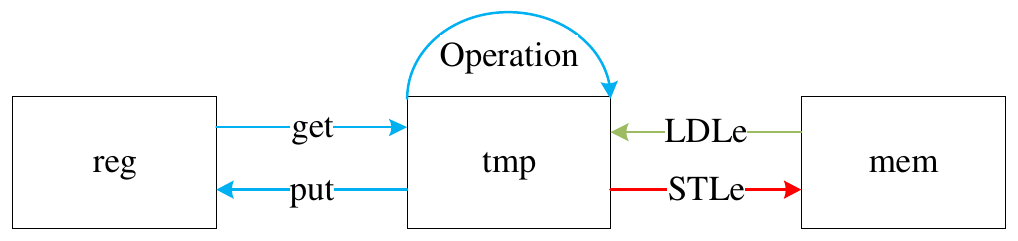}
	\caption{ VEX IR data transfer model}
	\label{data_transfer_model}
\end{figure}

\textbf{Pointer dependence analysis.}
The overall flow of pointer dependency analysis is as follows: 
DEPA first records the variable that holds the pointer to the heap object 
when the heap object allocation operation is recognized. 
According to the different IR operations, it then completes the transfer of pointer dependencies.
The VEX IR data transfer model is shown in Figure \ref{data_transfer_model}. 
The arrows represent the various operations of VEX IR, 
and the box represents the various data objects of VEX IR. 
The GET operation passes the value in the register into a temporary variable, 
and PUT is the reverse operation. 
The LDLe operation is to pass a value in memory to a temporary variable, 
and the STLe operation is the reverse. 
The Operation represents an arithmetic operation on a temporary variable.

\begin{figure}[h]
	\centering
	\includegraphics[width=\linewidth]{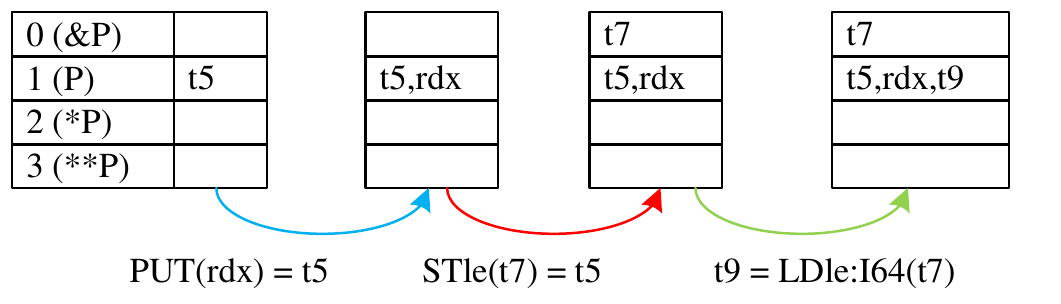}
	\caption{ VEX IR data transfer sample}
	\label{data_transfer_sample}
\end{figure}

Figure \ref{data_transfer_sample} shows a simple example to illustrate the process of pointer dependent passing. 
Create a pointer dependency table for the heap object pointer, 
assuming t5 is the variable that holds the heap object pointer at the start. 
The pointer dependency table is used to represent different levels of pointers, 
such as level 0 for the address of t5, level 2 for the data in memory that t5 points to, 
and level 3 for the data that level 2 point to. 
When different operations are encountered, the pointer dependency table is modified accordingly. 
The GET, PUT and OPERATION operations do not affect the pointer level, 
while the LDLe instruction will raise the pointer level and the STLe will lower the pointer level.

The target address of the write memory is analyzed through the pointer dependency table. 
If the target address is at level 1 of the dependency table, it is the object write operation. 
If it is at level 2 or 3 of the dependency, it is the dereference of the internal pointer of the object. 
Use the same method to analyze the destination address of the jump instruction. 
Finally, we can get the primitive-crucial-behaviors corresponding to the input, 
and then add the label representing the primitive-crucial-behaviors for the input to get the Tagged Inputs.

\subsection{Fuzzing Based on Generation}

For interactive programs, the input sequence with specific rules is needed to complete meta-operation. 
We use  generation-based fuzzing to generate more inputs that can explore deep code in the program 
and further trigger the exploit primitives.
As can be seen from the process of triggering the heap overflow vulnerability exploit primitive in Figure \ref{OOB_process} and the process of triggering the UAF exploit primitive in Figure \ref{UAF_process}, 
the triggering of the exploit primitive is related to the primitive-crucial-behavior and has a certain pattern. 
By analyzing program primitive-crucial-behaviors, 
we can get the primitive-crucial-behaviors corresponding to different types of input of interactive programs. 
Therefore, a template for generating input can be constructed by using the trigger mode of the primitive 
and the corresponding relationship between input types and primitive-crucial-behaviors. 
Then, in the Fuzzing stage, new inputs can be generated according to the template 
for Fuzzing to determine the input that can trigger the exploit primitive.
Legal inputs to an interactive program can be thought of as a formal language generated by 
a context-free grammar G =\{N, T, R, S\}, 
and an input that conforms to a particular template is equivalent to adding a corresponding constraint to R. 
More details about grammar G are as follows:

\begin{itemize}
	\item N is a finite set of non-terminators. Non-terminators can be thought as intermediate states in language specifications.
	\item T is a finite set of terminators. N and T do not intersect.
	\item R is a finite set containing generating rules of the form a $\rightarrow \alpha$ , where a $\in$ N, $\alpha$ $\in$ (T $\cup$ N)*.
	\item S $\in$ N is a non-terminator and a start symbol. Every word generated by context-free grammar needs to be derivable from S.
\end{itemize}

\subsection{Concolic Execution}

\begin{figure}[t]
	\centering
	\includegraphics[width=\linewidth]{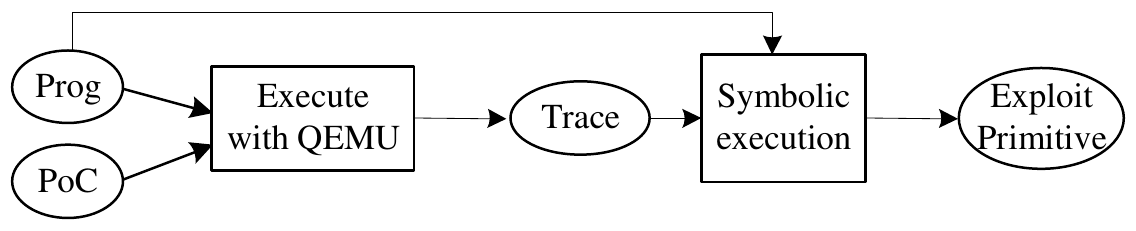}
	\caption{The process of exploit primitive determination}
	\label{exploit_primitive_determin}
\end{figure}

The POC that causes the program to crash can be retrieved from the fuzzing procedure. 
However, these inputs are not necessarily the ones that trigger the exploit primitives and need to be further verified. 
Figure \ref{exploit_primitive_determin} shows the process of exploit primitive determination.

The input of the process includes the target binary program Prog and the POC. 
Its output is an exploit primitive if the POC lead program to be exploitable.
The target program with POC is executed through the QEMU concrete execution module, 
and at the same time the execution path of the program is collected. 
Then DEPA execute the target program symbolically according to the specific path. 
When the execution reaches the crash state, DEPA judges whether the crash state is an exploit primitive. 

\subsection{Implementation}

The primitive-crucial-behavior analysis of the program is based on static analysis of angr \cite{angr}. 
Firstly, the control flow graph (CFG) of the program is generated using angr, 
and then the specific execution path of the program is obtained through the concrete execution of QEMU \cite{QEMU}. The IR of the basic block corresponding to the specific path is extracted from the control flow graph, 
and static analysis of the intermediate representation is carried out to determine the primitive-crucial-behavior. 
Generation-based fuzzing is implemented based on the Nautilus \cite{NAUTILUS}.
We modify the seed-generation syntax of Nautilus for different models.
The concolic execution is built on top of REX \cite{REX}.

\section{Evaluation}

We implemented a prototype of DEPA based on the binary analysis engine angr \cite{angr}.
Our experiment was run on Ubuntu Linux 16.04 system with hardware configuration 
including Intel i7 4th CPU and 8G memory. 
For fair comparison with Revery \cite{REVERY}, 
we evaluated DEPA against a set of CTF (capture the flag) programs. 
To demonstrate DEPA's effectiveness, we evaluate it to answer three research questions.

\begin{description}
\item[RQ1] Is the primitive-crucial-behavior analysis effective?
\item[RQ2] Is our approach superior to the state-of-the-art technology?
\item[RQ3] Is template-based fuzzing more effective than that without template?
\end{description}

We answer above questions in the following sections.

\subsection{Primitive-crucial-behavior analysis}

To verify the effectiveness of primitive-crucial-behavior analysis,
we ran DEPA on ten CTF programs and the results are shown in Table \ref{key_behaviors}, 
where the 2nd column means the binary size for each program,  
value $1$ in columns 3 -6  represents that the program contains the corresponding primitive-crucial-behaviors,  $0$ otherwise. 
The experimental results show that the primitive-crucial-behavior analysis module can successfully identify all the primitive-crucial-behaviors in these subjects. 
Since there is no ground truth to be referenced to verify the experimental results, 
we verify the correctness of the experimental results by manually analyzing the binaries. 
Furthermore, although we analyzed ten CTF subjects, they include dozens of key behaviors.
This demonstrates the effectiveness of our method for the analysis of primitive-crucial-behavior.

\begin{table*}
	\renewcommand\arraystretch{1.5} 
	\centering
	\caption{CTF programs and primitive-crucial-behaviors in them}
	\label{key_behaviors}
	\begin{tabular}{lrcccc}
		\hline
		Program& Size(KB) &allocation&release&deref writes&deref jump\\
		\hline
		main&20 &1&1&1&0\\
		marimo&15 &1&0&1&0\\
		simplememopad&9 &1&0&1&0\\
		woo2&14 &1&1&0&1\\
		woO2\_fixed&14 &1&1&0&1\\
		shop2&20 &1&1&0&1\\
		babyheap&10 &1&1&1&0\\
		b00ks&11 &1&0&1&0\\
		ezhp&14 &1&0&1&0\\
		note1&11 &1&0&1&0\\
		multi-heap& 9 &2&1&1&0\\
		
		\hline& 
	\end{tabular}
\end{table*}

\subsection{Compare to state-of-the-art technique}

In this section, we evaluate the effectiveness of DEPA by comparing it  with Revery 
which represents the most advanced solutions for exploit generation for Linux binaries. 
The experiment result is shown in Table \ref{compare_revery}. 
We used the experiment datasets from Revery, and the types of these vulnerabilities include UAF, heap overflow and off-by-one.  Revery fail to determine exploit primitive for $simplemempad$, whereas our method gets an arbitrary write primitive. The reason is that Revery cannot identify overflow vulnerabilities inside heap objects, and our tools do not presuppose vulnerability identification. It shows that our tool outperforms Revery in discovering exploit primitives for heap-based vulnerabilities.

For program $multi-heap$, our tool and Revery both failed to generate exploit primitives. The major reason is that $multi-heap$ is a multithreaded application that is still difficult to detect vulnerabilities for symbolic execution. This is out of scope of DEPA.

\begin{table*}
\renewcommand\arraystretch{1.5} 
\centering
	\caption{Determining exploit primitives using REVERY and DEPA}
	\label{compare_revery}
	\begin{tabular}{lccc}
		\hline
		Program&Vul Type&REVERY&DEPA\\
		\hline
		main&UAF&mem write&arbitrary write\\
		marimo&Heap Overflow&mem write&arbitrary write\\
		simplememopad&Heap Overflow&--&arbitrary write\\
		woO2&UAF&EIP hijack&arbitrary jump\\
		woO2\_fixed&UAF&EIP hijack&arbitrary jump\\
		shop2&UAF&EIP hijack&arbitrary jump\\
		babyheap&UAF&mem write&arbitrary write\\
		b00ks&off-by-one&mem write&arbitrary write\\
		ezhp&Heap Overflow&mem write&arbitrary write\\
		note1&Heap Overflow&mem write&arbitrary write\\
		multiheap& UAF &--&--\\
		\hline
	\end{tabular}
\end{table*}

\subsection{Template-based Fuzzing}

To verify whether template-based fuzzing is effective, 
the Fuzzing was conducted for 2 hours under the case with the template and without the template, respectively. 
Table \ref{Fuzzing_result} shows the result of fuzzing under different conditions. It can be seen that using templates is more effective in finding the exploit primitive. Specifically, with the help of templates, on one hand, DEPA produced fewer useless crashes that do not contain exploit primitives, and on the other hand, DEPA can find more exploit primitives.

\begin{table*}
\renewcommand\arraystretch{1.5} 
\centering
	\caption{Fuzzing results in different conditions}
	\label{Fuzzing_result}
	\begin{tabular}{lcccc}
  \hline
		Program&Crashes&Exploit primitives&Crashes&Exploit primitives\\
		 &w/o template&w/o template&with template&with template\\
     \hline
		main&11&0&4&1\\
		marimo&6&1&5&1\\
		simplememopad&4&1&9&1\\
		woo2&12&1&8&1\\
		woO2\_fixed&7&1&6&1\\
		shop2&10&1&12&1\\
		babyheap&5&0&10&1\\
		b00ks&15&1&10&1\\
		ezhp&8&1&7&1\\
		note1&10&1&6&1\\
    \hline
		Average&8.8&0.8&7.7&1.0\\
    \hline
	\end{tabular}
\end{table*}

\section{Case Studies}

DEPA successfully determined exploit primitives for 10 programs.
In this section, we investigate these programs and analyze why our solution DEPA succeeded. 

DEPA finds exploit primitives directly through customized Fuzzing and does not rely on vulnerability identification, 
so it was able to determine primitives for 10 programs, including $simplemempad$ which Revery cannot deal with. 

$Simplememopad$ is a typical interactive program for managing notes. 
It mainly contains three operations: $write\_note$, $edit\_note$ and $delete\_note$. 
Notes are organized by linked lists.
The $edit\_note$ operation contains an inner heap object overflow vulnerability 
that overwrites the $next\_pointer$ field in the object. 
The delete note operation contains an unsafe unlink operation, 
so an arbitrary address write primitive may be triggered.

DEPA first analyzed the PCBs of $write\_note$, $edit\_note$ and $delete\_note$ through $PCB\ Analysis$ module. 
The template for Fuzzing is then constructed based on the exploit primitive model and the relationship between PCBs and operations. 
The target program is then Fuzzing against the Fuzzing template to get crashes that may trigger the exploit primitive. 
In fact, a Fuzzing template is a sequence of operations, 
and the Fuzzing process is mainly a mutation of the specific parameters of the operation. 
Finally, the $Concolic\ Execution$ module analyzes crashes to find those that trigger the exploit primitive.

\section{Discussion}

Automatic exploit generation for general programs are very challenging or even impossible, 
we target at interactive programs in this paper. 
The reason behind it is that because interactive programs usually involve a large of complex conditions or infinite loops, which makes it hard to trigger stateful vulnerabilities in them. 
DEPA tackles the issue by leveraging generation-based fuzzing, 
which works better for interactive programs whose input follows a certain pattern.

We evaluated DEPA against eleven CTF programs instead of real-world programs, 
because the constraints solving ability of the binary analysis engine angr \cite{angr}  is not enough for complex programs. 
We leave it for future work.

\section{Related Work}

In this section, we discuss related work in the area of automatic exploit generation.
In 2011, Thanassis Avgerinos et al., for the first time, realized the Automatic end-to-end system AEG from source code vulnerability discovery to vulnerability exploit \cite{AEG}.
However, it can only exploit stack overflow vulnerability and format string vulnerability without considering any defense mechanism.
Both the CRAX \cite{CRAX} method and AEG detect vulnerabilities through symbolic execution, and use the test cases generated in the previous step to gather run-time information from the concrete execution. CRAX uses concolic execution directly to explore paths with vulnerability, to filter out complex and inaccessible function that have no effect on exploit generation, to reduce the overhead of the constraint solver (SMT). Finally, by monitoring the symbolic pointer and the symbolic program counter, CRAX determine the exploit primitive. CRAX abstracts exploits from different perspectives, so it can identify exploits of multiple vulnerability types. 

In 2012, Sang Kil Cha et al. implemented Mayhem \cite{Mayhem}, a tool for automatically finding exploitable bugs in binary files. Mayhem introduced a new concolic execution scheme that combines the advantages of existing symbol execution techniques (online and offline) into one system. Index-based memory modeling was also introduced, a technique that allowed Mayhem to find more exploitable bugs at the binary level. The authors mined and exploited 29 applications using Mayhem, which supports stack overflows and format string exploits.

In 2016, Yan Shoshitaishvili et al. implemented a variety of binary analysis techniques in a unified framework, angr \cite{angr}, which was used to automatically identify and exploit vulnerabilities in binary files. 
It provides convenience for the subsequent binary analysis and vulnerability exploitation. 
The authors of angr have also developed a vulnerability exploitation engine called REX \cite{REX} based on angr.
Given an application with vulnerability, REX although has an ability to reproduce the Crash path and analyze the register state and memory layout at the time of Crash to determine the exploitability of Crash and automatically generate exploits, it currently only supports simple buffer overflows on the stack.
To solve the problem of read exceptions causing the binary program to crash and thus hindering automatic exploit generation, Jiang et al. \cite{JIANG} proposed an exploitability analysis method based on automatic exception suppression to keep the program running when a read exception error that can crash the program is encountered.
By the way, they convert a read exception that cannot be exploited for existing automatic tools  into a write exception or a execute exception to effectively assess the exploitability of the read exception for binary vulnerability.

As we all know, it is an arduous task to automatically generate exploit for heap overflows. 
Therefore, Liang He et al. first implemented a prototype HCSIFTER \cite{HCSIFTER} to automatically assess the exploitability of heap overflows to identify whether a heap overflow vulnerability is exploitable or not rather than generate a working exploit. They aim to rank the severity of heap overflow vulnerabilities and thus prioritize resources.
HCSIFTER can accurately find the locations of the heap overflows through dynamic taint analysis \cite{TAINT}.  
And it uses dynamic heap data recovery and exploit point detection methods to extract necessary information to quantify the severity of the heap overflows based on attack metrics and feasibility metrics.

In order to find exploit primitives for heap-based metadata vulnerabilities to accelerate the process of generating working exploits, Repel et al. \cite{REPEL} presented an approach to detect reusable attack patterns against heap managers. Eckert et al. \cite{ECKERT} utilized model checking and symbolic execution technologies for analyzing the exploitability of heap implementations and finding the weaknesses in heap metadata defenses to achieve heap metadata attacks. Deng et al. \cite{DENG} proposed a pattern-based software testing framework for simulating human exploitation behavior at machine level in order to assess the exploitability of metadata corruption vulnerabilities. With the help of human security expertise, they make use of the heap layout serialization method to construct exploit patterns to guide the migration of memory state and then drive exploitation for heap metadata vulnerabilities at a small cost.

In 2018, Wang et al. proposed a solution, Revery \cite{REVERY}, which aims at automatic generation of heap overflow vulnerability exploitation. It can search exploitable states in divergent paths rather than just crash paths and adopts the fuzzing for memory layout. It addresses the challenges faced by existing AEG solutions such as exploit derivability, symbolic execution bottlenecks, and heap-based vulnerability exploitation. It can trigger a vulnerability and exploitability state for most vulnerable applications. It can also successfully generate working exploits for certain vulnerabilities. However, Revery cannot identify the internal buffer overflow vulnerability of the heap object whereas DEPA dose. 
To pave the way for automated exploit generation in the field of heap vulnerabilities, Wang et al. researched automated heap layout manipulation and presented a solution MAZE \cite{MAZE}, which used a $Dig  \delta  Fill$ algorithm to generate target heap layout . Although it is very efficient and effective compared to existing solutions, a POC is required to perform MAZE.

To facilitate the exploitability of use-after-free vulnerabilities in the Linux kernel, Xu et al. presented a memory collision attack approach to manipulate data in uninitialized memory regions \cite{Collision}. They proposed two collision attack strategies, an object-based attacking and a physmap-based attacking,  for increasing the success rate of the attack.  Wu et al. further developed FUZE \cite{FUZE} that can find a useful exploitation primitive to toward assisting to augment a security analyst with the ability to automatic exploit generation for use-after-free in the Linux kernel.  FUZE uses kernel fuzzing technique to find out the context of different kernel panic and then utilizes symbolic execution along with dynamic tracing to explore exploitability under different contexts.

To escalate the ability of automatic exploit generation for Linux kernel vulnerabilities, Yueqi Chen et al. proposed the Slake \cite{SLAKE} method to determine the data objects that are useful for kernel exploit and how to construct the desired memory layout to facilitate the evaluation of Linux kernel vulnerability exploitability.  Besides, Wei Wu et al. proposed KEPLER \cite{KEPLER} to facilitate evaluation of control-flow hijacking primitives in the Linux kernel. For kernel out-of-bounds write vulnerabilities, Weiteng Chen et al. proposed KOOBE \cite{KOOBE}, which firstly analyzes POC with symbol execution, summarizes POC's capabilities using a capability-guided fuzzing solution, and then evaluates its exploitability according to POC's capabilities and target objects.

\section{Conclusion}

In this paper, we propose a solution DEPA to determine exploit primitives of vulnerabilities in interactive programs, which  is based on the analysis of primitive-crucial-behaviors. 
Our approach first analyzes the program's primitive-crucial-behaviors under different types of input, 
then constructs fuzzing template according to the primitive-crucial-behavior model that triggers the exploit primitive, next performs fuzzing with the above template, 
and finally determines through symbolic execution whether the generated input causing crash is an effective exploit primitive. 
We evaluate DEPA on ten real-world CTF programs and experiment results show that DEPA can identify primitive-crucial-behavior accurately and determine exploit primitives effectively. 
In addition, DEPA is superior to the state-of-the-art tool Revery in determining exploit primitives for the heap internal overflow vulnerability.

\bibliographystyle{IEEEtran}
\bibliography{samplepaper}

\end{document}